%% file: main.tex
\pdfoutput=1

\documentclass[12pt,a4paper]{article}

\usepackage{ifthen} 
\newboolean{pdflatex}
\setboolean{pdflatex}{true} 

\newboolean{articletitles}
\setboolean{articletitles}{true} 

\newboolean{uprightparticles}
\setboolean{uprightparticles}{false} 

\def\paperauthors{LHCb collaboration} 
\def\paperasciititle{Measurement of eta_c production cross-section in pp collisions at sqrt(s) = 13 TeV} 
\def\papertitle{Measurement of the $\etac(1S)$ production cross-section in \proton\proton collisions at $\sqrt{s} = 13$\tev} 
\def\paperkeywords{{High Energy Physics}, {LHCb}}
\def\papercopyright{\the\year\ CERN for the benefit of the LHCb collaboration} 
\def\paperlicence{CC-BY-4.0 licence}
\def\paperlicenceurl{https://creativecommons.org/licenses/by/4.0/}

\input{preamble}
\usepackage{subfigure}
\usepackage{longtable} 

\usepackage{framed,color}
\usepackage{rotating}
\definecolor{shadecolor}{rgb}{1,1,1}
\usepackage{amsmath,amsfonts,amsthm,bm}
\usepackage{tabularx}
\usepackage{url}
\usepackage[all]{nowidow}

\begin{document}

\renewcommand{\thefootnote}{\fnsymbol{footnote}}
\setcounter{footnote}{1}

\newcommand{\tzfit}{$t_z$-fit technique \xspace}
\newcommand{\tzcut}{separation technique \xspace}
\newcommand{\ppbar}{\proton\antiproton}
\newcommand{\JpsiToPpbar}{\decay{\jpsi}{\ppbar}}
\newcommand{\JpsiToPpbarPiz}{\decay{\jpsi}{\ppbar\piz}}
\newcommand{\EtacToPpbar}{\decay{\etac}{\ppbar}}
\newcommand{\bToEtacX}{\decay{\bquark}{\etac X}}
\newcommand{\bToJpsiX}{\decay{\bquark}{\jpsi X}} 
\newcommand{\etacPromptRelativeXsec}{1.69 \pm 
									 0.15 \pm 
									 0.10 \pm 
									 0.18}
\newcommand{\etacPromptAbsoluteXsec}{1.26 \pm
									 0.11\pm
									 0.08\pm
									 0.14 \mub}
\newcommand{\jpsiPromptAbsoluteXsec}{0.749 \pm
									 0.005 \pm
									 0.028 \pm
									 0.037 \mub} 
\newcommand{\etacSecondaryRelativeBR}{0.48 \pm 0.03 \pm 0.03 \pm 0.05} 
\newcommand{\etacSecondaryAbsoluteBR}{(5.51 \pm 0.32 \pm 0.29 \pm 0.77)\times 10^{-3}} 

\newcommand{\etacMassDiffTzFitSim}{111.2\pm1.1~\mev}
\newcommand{\jpsiMassTzFitSim}{3096.6\pm0.1~\mev}

\newcommand{\etacMassDiffTzCutInt}{112.7\pm0.8~\mev}
\newcommand{\jpsiMassTzCutInt}{3096.6\pm0.1~\mev}

\newcommand{\resoSim}{7.78\pm0.12}

\newcommand{\epsRatioJpsipppizJpsipp}{0.062\pm0.002}

\newcommand{\etacMassDiff}{113.0 \pm
					       0.7 \pm
						   0.1 \mev}

\newcommand{\etacMassDiffInBins}{113.22\pm0.67 \mev}

\newcommand{\etacMassDiffPDG}{113.5\pm0.5 \mev}
\newcommand{\jpsiMassPDG}{(3096.900\pm0.006) \mev}
\newcommand{\brEtacpp}{(1.50\pm0.16)\times10^{-3} }
\newcommand{\brJpsipp}{(2.120\pm0.029)\times10^{-3} }
\newcommand{\brJpsipppiz}{(1.19\pm0.08)\times10^{-3} }
\newcommand{\brRatioJpsipppizJpsipp}{0.56\pm0.04 }

\input{title-LHCb-PAPER}


\renewcommand{\thefootnote}{\arabic{footnote}}
\setcounter{footnote}{0}



\pagestyle{plain} 
\setcounter{page}{1}
\pagenumbering{arabic}


%

\input{introduction}

\input{body}

\input{tzcut}
\input{mass}
\input{summary}

\input{acknowledgements}

\input{appendix}


\addcontentsline{toc}{section}{References}
\bibliographystyle{LHCb}
\bibliography{main,standard,LHCb-PAPER,LHCb-CONF,LHCb-DP,LHCb-TDR}

\newpage
\input{LHCb_Authorship_flat_11-Jun-2019}



\end{document}

%% file: preamble.tex
\usepackage[top=1in, bottom=1.25in, left=1in, right=1in]{geometry}

\usepackage{microtype}
\usepackage{lineno}  
\usepackage{xspace} 
\usepackage{caption} 

\usepackage{graphicx}  
\usepackage{color}
\usepackage{colortbl}
\graphicspath{{./figs/}} 
\DeclareGraphicsExtensions{.pdf,.PDF,png,.PNG}

\usepackage{amsmath} 
\usepackage{amssymb}
\usepackage{amsfonts}
\usepackage{upgreek} 

\newcommand*\patchAmsMathEnvironmentForLineno[1]{%
\expandafter\let\csname old#1\expandafter\endcsname\csname #1\endcsname
\expandafter\let\csname oldend#1\expandafter\endcsname\csname
end#1\endcsname
 \renewenvironment{#1}%
   {\linenomath\csname old#1\endcsname}%
   {\csname oldend#1\endcsname\endlinenomath}%
}
\newcommand*\patchBothAmsMathEnvironmentsForLineno[1]{%
  \patchAmsMathEnvironmentForLineno{#1}%
  \patchAmsMathEnvironmentForLineno{#1*}%
}
\AtBeginDocument{%
\patchBothAmsMathEnvironmentsForLineno{equation}%
\patchBothAmsMathEnvironmentsForLineno{align}%
\patchBothAmsMathEnvironmentsForLineno{flalign}%
\patchBothAmsMathEnvironmentsForLineno{alignat}%
\patchBothAmsMathEnvironmentsForLineno{gather}%
\patchBothAmsMathEnvironmentsForLineno{multline}%
\patchBothAmsMathEnvironmentsForLineno{eqnarray}%
}

\usepackage{hyperxmp}

\usepackage[pdftex,
            pdfauthor={\paperauthors},
            pdftitle={\paperasciititle},
            pdfkeywords={\paperkeywords},
            pdfcopyright={Copyright (C) \papercopyright},
            pdflicenseurl={\paperlicenceurl}]{hyperref}

\usepackage[colorinlistoftodos,textsize=scriptsize]{todonotes}

\usepackage[all]{hypcap} 

\input{lhcb-symbols-def} 

\usepackage{cite} 
\usepackage{mciteplus}

%% file: lhcb-symbols-def.tex
\usepackage{xspace} 
\usepackage{upgreek}

\def\lhcb   {\mbox{LHCb}\xspace}

\def\babar  {\mbox{BaBar}\xspace}

\def\lhc    {\mbox{LHC}\xspace}
\def\lep    {\mbox{LEP}\xspace}
\def\tevatron {Tevatron\xspace}

\def\rich   {RICH\xspace}

\def\MagUp {\mbox{\em Mag\kern -0.05em Up}\xspace}

\ifthenelse{\boolean{uprightparticles}}%
{

 \def\Peta        {\ensuremath{\upeta}\xspace}

 \def\Pmu         {\ensuremath{\upmu}\xspace}

 \def\Ppi         {\ensuremath{\uppi}\xspace}

 \def\Pchi        {\ensuremath{\upchi}\xspace}                 
 \def\Ppsi        {\ensuremath{\uppsi}\xspace}

 \def\PDelta      {\ensuremath{\Delta}\xspace}                 
 \def\PXi         {\ensuremath{\Xi}\xspace}                 
 \def\PLambda     {\ensuremath{\Lambda}\xspace}                 
 \def\PSigma      {\ensuremath{\Sigma}\xspace}                 
 \def\POmega      {\ensuremath{\Omega}\xspace}                 
 \def\PUpsilon    {\ensuremath{\Upsilon}\xspace}

 \def\PB      {\ensuremath{\mathrm{B}}\xspace}                 
                  
 \def\PD      {\ensuremath{\mathrm{D}}\xspace}

 \def\PJ      {\ensuremath{\mathrm{J}}\xspace}                 
 \def\PK      {\ensuremath{\mathrm{K}}\xspace}

 \def\Pb      {\ensuremath{\mathrm{b}}\xspace}                 
 \def\Pc      {\ensuremath{\mathrm{c}}\xspace}

 \def\Pi      {\ensuremath{\mathrm{i}}\xspace}

 \def\Pp      {\ensuremath{\mathrm{p}}\xspace}

 \def\Ps      {\ensuremath{\mathrm{s}}\xspace}

 \def\thebaroffset{0.0em}
}
{

 \def\Peta        {\ensuremath{\eta}\xspace}

 \def\Pmu         {\ensuremath{\mu}\xspace}

 \def\Ppi         {\ensuremath{\pi}\xspace}

 \def\Pchi        {\ensuremath{\chi}\xspace}                 
 \def\Ppsi        {\ensuremath{\psi}\xspace}                 
                  
 \mathchardef\PDelta="7101
 \mathchardef\PXi="7104
 \mathchardef\PLambda="7103
 \mathchardef\PSigma="7106
 \mathchardef\POmega="710A
 \mathchardef\PUpsilon="7107
                  
 \def\PB      {\ensuremath{B}\xspace}                 
                  
 \def\PD      {\ensuremath{D}\xspace}

 \def\PJ      {\ensuremath{J}\xspace}                 
 \def\PK      {\ensuremath{K}\xspace}

 \def\Pb      {\ensuremath{b}\xspace}                 
 \def\Pc      {\ensuremath{c}\xspace}

 \def\Pi      {\ensuremath{i}\xspace}

 \def\Pp      {\ensuremath{p}\xspace}

 \def\Ps      {\ensuremath{s}\xspace}

 \def\thebaroffset{0.18em}
}
\newcommand{\offsetoverline}[2][\thebaroffset]{\kern #1\overline{\kern -#1 #2}}%

\makeatletter
\ifcase \@ptsize \relax
  \newcommand{\miniscule}{\@setfontsize\miniscule{4}{5}}
\or
  \newcommand{\miniscule}{\@setfontsize\miniscule{5}{6}}
\or
  \newcommand{\miniscule}{\@setfontsize\miniscule{5}{6}}
\fi
\makeatother

\DeclareRobustCommand{\optbar}[1]{\shortstack{{\miniscule (\rule[.5ex]{1.25em}{.18mm})}
  \\ [-.7ex] $#1$}}

\def\mup        {{\ensuremath{\Pmu^+}}\xspace}
\def\mun        {{\ensuremath{\Pmu^-}}\xspace} 

\def\squark    {{\ensuremath{\Ps}}\xspace}

\def\cquark    {{\ensuremath{\Pc}}\xspace}

\def\bquark    {{\ensuremath{\Pb}}\xspace}

\def\pion   {{\ensuremath{\Ppi}}\xspace}
\def\piz    {{\ensuremath{\pion^0}}\xspace}

\def\kaon    {{\ensuremath{\PK}}\xspace}

\def\KorKbar {\kern \thebaroffset\optbar{\kern -\thebaroffset \PK}{}\xspace}

\def\Kp      {{\ensuremath{\kaon^+}}\xspace}

\def\KS      {{\ensuremath{\kaon^0_{\mathrm{S}}}}\xspace}

\def\DorDbar {\kern \thebaroffset\optbar{\kern -\thebaroffset \PD}\xspace}

\def\B       {{\ensuremath{\PB}}\xspace}

\def\BorBbar {\kern \thebaroffset\optbar{\kern -\thebaroffset \PB}\xspace}
\def\Bz      {{\ensuremath{\B^0}}\xspace}

\def\Bd      {{\ensuremath{\B^0}}\xspace}

\def\BdorBdbar {\kern \thebaroffset\optbar{\kern -\thebaroffset \Bd}\xspace}
\def\Bu      {{\ensuremath{\B^+}}\xspace}

\def\Bp      {{\ensuremath{\Bu}}\xspace}

\def\Bs      {{\ensuremath{\B^0_\squark}}\xspace}

\def\BsorBsbar {\kern \thebaroffset\optbar{\kern -\thebaroffset \Bs}\xspace}

\def\Bcp     {{\ensuremath{\B_\cquark^+}}\xspace}

\def\jpsi     {{\ensuremath{{\PJ\mskip -3mu/\mskip -2mu\Ppsi\mskip 2mu}}}\xspace}

\def\etac     {{\ensuremath{\Peta_\cquark}}\xspace}

\def\chiczero {{\ensuremath{\Pchi_{\cquark 0}}}\xspace}
\def\chicone  {{\ensuremath{\Pchi_{\cquark 1}}}\xspace}
\def\chictwo  {{\ensuremath{\Pchi_{\cquark 2}}}\xspace}

\def\Y#1S{\ensuremath{\PUpsilon{(#1S)}}\xspace}

\def\proton      {{\ensuremath{\Pp}}\xspace}
\def\antiproton  {{\ensuremath{\overline \proton}}\xspace}

\def\LorLbar     {\kern \thebaroffset\optbar{\kern -\thebaroffset \PLambda}\xspace}

\def\BF         {{\ensuremath{\mathcal{B}}}\xspace}
\def\BR         {\BF}

\newcommand{\decay}[2]{\ensuremath{#1\!\to #2}\xspace} 

\def\to                 {\ensuremath{\rightarrow}\xspace}

\def\AT#1     {\ensuremath{A_{\mathrm{T}}^{#1}}\xspace}           

\def\C#1      {\ensuremath{\mathcal{C}_{#1}}\xspace}                       
\def\Cp#1     {\ensuremath{\mathcal{C}_{#1}^{'}}\xspace}                    
\def\Ceff#1   {\ensuremath{\mathcal{C}_{#1}^{\mathrm{(eff)}}}\xspace}        
\def\Cpeff#1  {\ensuremath{\mathcal{C}_{#1}^{'\mathrm{(eff)}}}\xspace}       
\def\Ope#1    {\ensuremath{\mathcal{O}_{#1}}\xspace}                       
\def\Opep#1   {\ensuremath{\mathcal{O}_{#1}^{'}}\xspace}                    

\newcommand{\nospaceunit}[1]{\ensuremath{\text{#1}}}       
\newcommand{\aunit}[1]{\ensuremath{\text{\,#1}}}       

\newcommand{\tev}{\aunit{Te\kern -0.1em V}\xspace}
\newcommand{\gev}{\aunit{Ge\kern -0.1em V}\xspace}
\newcommand{\mev}{\aunit{Me\kern -0.1em V}\xspace}
\newcommand{\kev}{\aunit{ke\kern -0.1em V}\xspace}
\newcommand{\ev}{\aunit{e\kern -0.1em V}\xspace}
\newcommand{\mevc}{\ensuremath{\aunit{Me\kern -0.1em V\!/}c}\xspace}
\newcommand{\gevc}{\ensuremath{\aunit{Ge\kern -0.1em V\!/}c}\xspace}
\newcommand{\mevcc}{\ensuremath{\aunit{Me\kern -0.1em V\!/}c^2}\xspace}
\newcommand{\gevcc}{\ensuremath{\aunit{Ge\kern -0.1em V\!/}c^2}\xspace}

\def\mum  {\ensuremath{\,\upmu\nospaceunit{m}}\xspace}

\def\mub{\ensuremath{\,\upmu\nospaceunit{b}}\xspace}
\def\nb {\aunit{nb}\xspace}

\def\fb   {\ensuremath{\aunit{fb}}\xspace}
\def\invfb   {\ensuremath{\fb^{-1}}\xspace}

\def\fs   {\aunit{fs}}

\newcommand{\chisq}{\ensuremath{\chi^2}\xspace}

\def\gsim{{~\raise.15em\hbox{$>$}\kern-.85em
          \lower.35em\hbox{$\sim$}~}\xspace}
\def\lsim{{~\raise.15em\hbox{$<$}\kern-.85em
          \lower.35em\hbox{$\sim$}~}\xspace}

\def\sqs   {\ensuremath{\protect\sqrt{s}}\xspace}

\def\pt         {\ensuremath{p_{\mathrm{T}}}\xspace}

\def\ptot       {\ensuremath{p}\xspace}

\def\evtgen     {\mbox{\textsc{EvtGen}}\xspace}

\def\geant      {\mbox{\textsc{Geant4}}\xspace}

\def\photos     {\mbox{\textsc{Photos}}\xspace}

\def\pythia     {\mbox{\textsc{Pythia}}\xspace}

\def\tell1  {TELL1\xspace}
\def\ukl1   {UKL1\xspace}



%% file: title-LHCb-PAPER.tex

\begin{titlepage}
\pagenumbering{roman}

\vspace*{-1.5cm}
\centerline{\large EUROPEAN ORGANIZATION FOR NUCLEAR RESEARCH (CERN)}
\vspace*{1.5cm}
\noindent
\begin{tabular*}{\linewidth}{lc@{\extracolsep{\fill}}r@{\extracolsep{0pt}}}
\ifthenelse{\boolean{pdflatex}}
{\vspace*{-1.5cm}\mbox{\!\!\!\includegraphics[width=.14\textwidth]{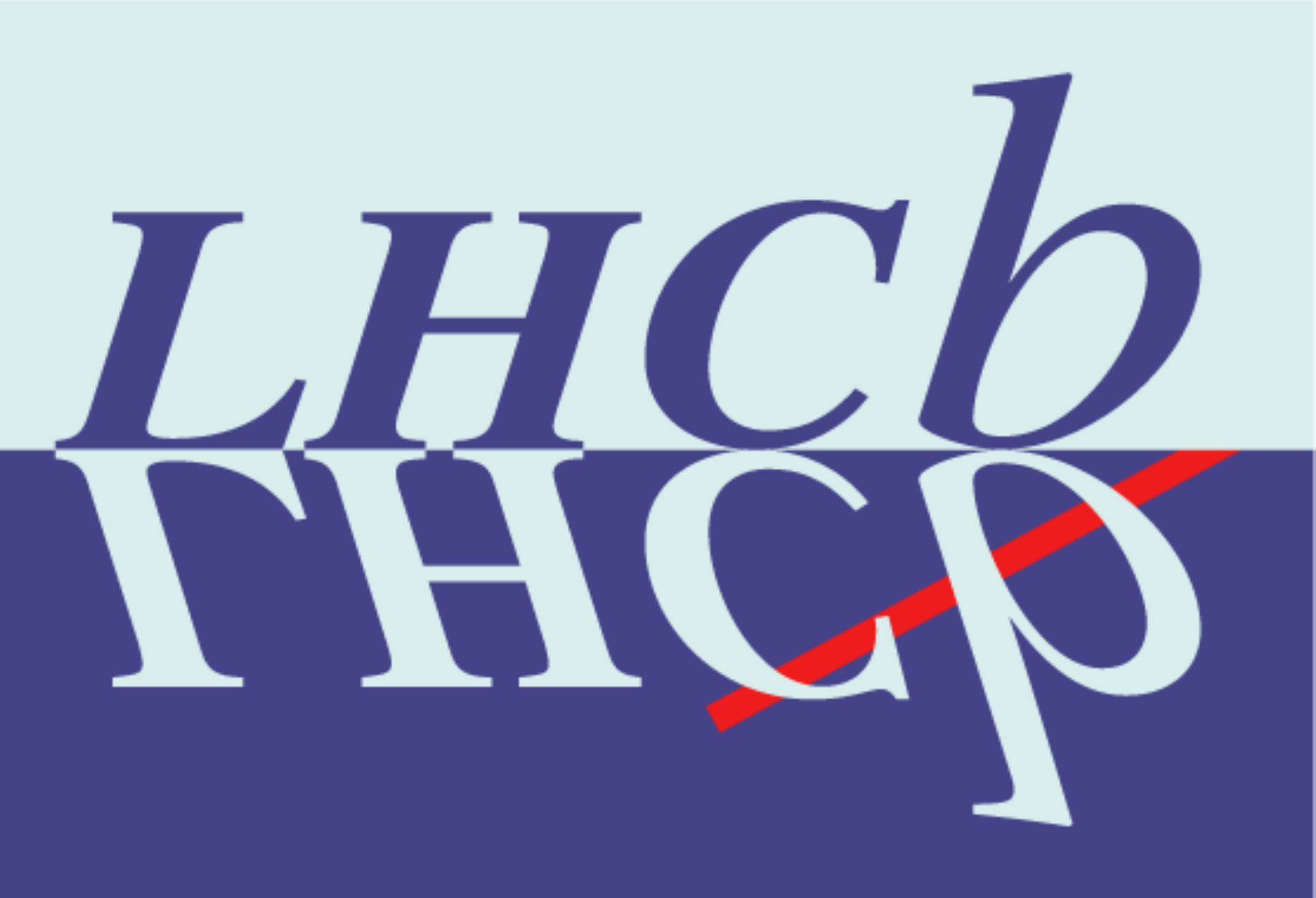}} & &}%
{\vspace*{-1.2cm}\mbox{\!\!\!\includegraphics[width=.12\textwidth]{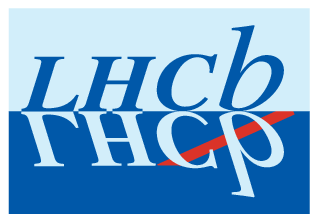}} & &}%
\\
 & & CERN-EP-2019-214 \\  
 & & LHCb-PAPER-2019-024 \\  
 & & 5 March 2020 \\ 
 & & \\
\end{tabular*}

\vspace*{2.0cm}

{\normalfont\bfseries\boldmath\huge
\begin{center}
  \papertitle 
\end{center}
}

\vspace*{1.0cm}

\begin{center}
\paperauthors\footnote{Authors are listed at the end of this paper.}
\end{center}

\vspace{\fill}

\begin{abstract}
  \noindent
Using a data sample corresponding to an integrated luminosity of 2.0 \invfb, collected by the \lhcb experiment, the production
of the $\etac(1S)$ state in \mbox{proton-proton} collisions at a centre-of-mass energy of $\sqs=13$\tev is studied in the rapidity range \mbox{$2.0 < y < 4.5$} and in the transverse momentum range \mbox{$6.5 < \pt < 14.0 \gev$}.
The cross-section for prompt production of $\etac(1S)$ mesons relative to that of the \jpsi meson is measured using the \ppbar decay mode and is found to be ${\sigma_{\etac(1S)}/\sigma_{\jpsi} = \etacPromptRelativeXsec}$. The quoted uncertainties are, in order, statistical, systematic and due to uncertainties on the branching fractions of the \JpsiToPpbar and \EtacToPpbar decays.
The prompt $\etac(1S)$  production cross-section is determined to be 
${\sigma_{\etac(1S)} = \etacPromptAbsoluteXsec}$,
where the last uncertainty includes that on the \jpsi meson cross-section. 
The ratio of the branching fractions of \bquark-hadron decays to the $\etac(1S)$ and \jpsi states is measured to be ${\BR_{\bToEtacX}/\BR_{\bToJpsiX} = \etacSecondaryRelativeBR}$,
where the last uncertainty is due to those on the branching fractions of the \JpsiToPpbar and \EtacToPpbar decays.
The difference between the \jpsi and $\etac(1S)$ masses is also determined to be $113.0 \pm 0.7 \pm 0.1 \mev$, which is the most precise single measurement of this quantity to date.
\end{abstract}
\vspace*{2.0cm}

\begin{center}
    Published in Eur. Phys. J. C80 (2020) 191
\end{center}

\vspace{\fill}

{\footnotesize 
\centerline{\copyright~\papercopyright. \href{\paperlicenceurl}{\paperlicence}.}}
\vspace*{2mm}

\end{titlepage}


\newpage
\setcounter{page}{2}
\mbox{~}
%
%
%
%

\cleardoublepage

%% file: introduction.tex
\newpage
\section{Introduction}
\label{sec:introduction}
Nonrelativistic quantum chromodynamics (NRQCD)~\cite{Bodwin:1994jh} is a powerful framework to describe the production from initial parton scattering (hadroproduction) of charmonium states with quantum numbers $J^{PC} = 1^{--}$, for instance the \jpsi meson, over a wide range of transverse momentum, \pt, and rapidity, $y$.
Nevertheless, it remains a challenge to provide a comprehensive theoretical description of measurements of the prompt production, comprising the hadroproduction and feed-down from excited resonant states, and the polarisation 
of \jpsi mesons at the \tevatron~\cite{Abe:1992ww} and the \lhc~\cite{LHCb-PAPER-2012-039,LHCb-PAPER-2013-059,LHCb-PAPER-2013-016,LHCb-PAPER-2015-037,Aad:2011sp,Aad:2015duc,Chatrchyan:2011kc,Sirunyan:2017qdw,Abelev:2012kr,Aamodt:2011gj,LHCb-PAPER-2013-008,LHCb-PAPER-2013-067,Chatrchyan:2013cla,Abelev:2011md,Abulencia:2007us} collision energies over the entire \pt range. 

A factorisation approach together with a heavy-quark spin symmetry assumption allows the simultaneous treatment of \jpsi and $\etac(1S)$ (the ground-level charmonium state with $J^{PC}=0^{-+}$ quantum numbers)\footnote{The $\etac(1S)$ meson is denoted as \etac throughout the rest of this paper.} production observables by imposing relations between their long-distance matrix elements (LDME)~\cite{Bodwin:1994jh}.
The \lhcb collaboration measured the prompt $\etac$  production cross-section in proton-proton collisions at centre-of-mass energies of $\sqs = 7$ and $8 \tev$~\cite{LHCb-PAPER-2014-029} to be below 
the predictions based on \jpsi prompt production data~\cite{Likhoded:2014fta,Butenschoen:2014dra,Han:2014jya,Zhang:2014ybe,Bodwin:2014gia}, 
which motivated several groups to revisit the theoretical approach~\cite{Feng:2015cba,Sun:2015pia,Likhoded:2015qyl,Gao:2016ihc,Faccioli:2017hym,Butenschoen:2017iks,Baranov:2016clx}.
A study of the $\etac$ prompt production at $\sqs = 13 \tev$ provides a further important test for theories predicting
the \jpsi and $\etac$ hadroproduction cross-sections and the \jpsi polarisation~\cite{Feng:2019zmn}.

The \lhcb collaboration has also measured the branching fractions of inclusive \mbox{\bquark-hadron} decays to $\etac$~\cite{LHCb-PAPER-2014-029} 
and to \chiczero, \chicone and \chictwo mesons~\cite{LHCb-PAPER-2017-007}. At the \lhc, the \bquark-hadron sample comprises a mixture of \Bp, \Bz, \Bs, \Bcp mesons and \bquark~baryons.\footnote{Charge conjugation is implied throughout the paper.}
A simultaneous study of the hadroproduction and production in inclusive \bquark-hadron decays of the charmonium states with linked 
LDMEs provides a unique test of basic NRQCD assumptions~\cite{LALpreprint}. Only marginal consistency was found between measurements and theoretical predictions at next-to-leading-order~\cite{Han:2014jya,Beneke:1998ks} for both prompt production and production in \bquark-hadron decays for the $\etac$ and \jpsi states.

Using a sample of ${\decay{\Bp}{\ppbar \Kp}}$ decays, the \lhcb collaboration has recently measured~\cite{LHCb-PAPER-2016-016} the mass difference of the \jpsi and $\etac$ states, $\Delta M_{\jpsi , \etac}$ to be 2.8 standard deviations smaller than the world-average value~\cite{PDG2018}.
A dataset of \bquark-hadron decays to the \etac meson can be analysed to measure the $\Delta M_{\jpsi , \etac}$ with improved precision.

This paper reports measurements of the \etac prompt production cross-section and branching fraction of \bquark-hadron inclusive decays to the \etac meson. A dedicated selection of \etac mesons produced in 
\bquark-hadron decays is developed to perform the most precise measurement of the $\Delta M_{\jpsi , \etac}$. Both \jpsi and \etac mesons are reconstructed via their decays to \ppbar.

%% file: body.tex
\section{The \lhcb detector and data sample} 
\label{sec:data}
The \lhcb detector~\cite{Alves:2008zz,LHCb-DP-2014-002} is a single-arm forward spectrometer covering 
the \mbox{pseudorapidity} range $2<\eta <5$, designed for the study of particles containing \bquark or \cquark quarks. 
The detector includes a high-precision tracking system consisting of a silicon-strip vertex detector surrounding the $pp$
interaction region, a large-area silicon-strip detector located upstream of a dipole magnet with a bending power of about
$4{\mathrm{\,Tm}}$, and three stations of silicon-strip detectors and straw drift tubes placed downstream of the magnet.
The tracking system provides a measurement of the momentum, \ptot, of charged particles with a relative uncertainty that varies 
from 0.5\% at low momentum to 1.0\% at 200\gev.\footnote{Natural units are used throughout the paper.}
The minimum distance of a track to a \proton\proton collision vertex (PV), the impact parameter (IP), is measured with a resolution of $(15+29/\pt)\mum$, where \pt is given in \gev.
Different types of charged hadrons are distinguished using information from two ring-imaging Cherenkov (\rich) detectors. 
Photons, electrons and hadrons are identified by a calorimeter system consisting of scintillating-pad and preshower detectors, 
an electromagnetic and a hadronic calorimeter. 
Muons are identified by a system composed of alternating layers of iron and multiwire proportional chambers. 
The online event selection is performed by a trigger, 
which consists of a hardware stage, based on information from the calorimeter and muon
systems, followed by a software stage, which applies a full event reconstruction.

The analysis is based on \proton\proton collision data recorded by the \lhcb experiment in 2015 and 2016
at a centre-of-mass energy of $13 \tev$, corresponding to an integrated luminosity of $2.0 \invfb$.
Events enriched in signal decays are selected by the hardware trigger based on the presence of a single deposit of high transverse energy in the calorimeter.
The trigger also specifically rejects high-multiplicity events, which produce excessive random combinations of tracks (combinatorial background).
The subsequent software trigger selects 
charged particles with a good track-fit quality and $\pt> 2\gev$. 
Proton candidates are identified using information from the \rich and tracking detectors. 
Pairs of oppositely charged proton candidates are required to form a good quality vertex and to have $\pt > 6.5\gev$.
The selection follows that of Ref.~\cite{LHCb-PAPER-2014-029}.
The signal selection of both prompt charmonia and charmonia from \bquark-hadron decays is largely performed at the trigger level.

For the measurement of the $\etac$ mass, a low-background data sample enriched in \bToEtacX decays is used. This sample is obtained using a software-trigger selection based on a multivariate algorithm that requires the presence of two, three or four charged tracks that form a common vertex and are inconsistent with originating from a PV~\cite{BBDT,LHCb-PROC-2015-018}.
Precise mass measurements require a momentum-scale calibration. 
The absolute scale is determined using $\Bp \to \jpsi \Kp$ decays with known particle masses as input~\cite{PDG2018}. 
Decays of $\jpsi \to \mup \mun$ are used to cross-calibrate a relative momentum scale between different data-taking periods~\cite{LHCb-PAPER-2012-048}.
The final calibration is checked with a variety of reconstructed quarkonia, \Bp and \KS meson decays.
No residual momentum bias is observed within the experimental resolution.

Samples of simulated events are used to model the effects of the detector acceptance and the imposed selection requirements.
In the simulation, $pp$ collisions are generated using \pythia~\cite{Sjostrand:2006za,Sjostrand:2007gs} 
with a specific \lhcb configuration~\cite{LHCb-PROC-2010-056}.  
Decays of hadronic particles are described by \evtgen~\cite{Lange:2001uf}, 
in which final-state radiation is generated using \photos~\cite{Golonka:2005pn}. 
The interaction of the generated particles with the detector material, and its response, are implemented 
using the \geant toolkit~\cite{Allison:2006ve, *Agostinelli:2002hh} as described in Ref.~\cite{LHCb-PROC-2011-006}.
A simulated sample of $\decay{\jpsi}{\ppbar\piz}$ decays is used to study the corresponding background contribution. The \etac and \jpsi decays are generated with uniform phase-space density, and the prompt \jpsi mesons are generated without polarisation. 
Inclusive \bquark-hadron decays are modelled using a combination of many exclusive final states based on measurements from the \B factories, \tevatron and \lhc experiments~\cite{Lange:2001uf}.

\section{Analysis technique} 
\label{sec:xsecDetermination}
In this analysis, \etac production is studied in a fiducial region of ${6.5 <\pt<14.0 \gev}$ and ${2.0<y<4.5}$.
The measurement of the differential production cross-section is performed as a function of the transverse momentum relative to that of the \jpsi meson~\cite{LHCb-PAPER-2015-037,PDG2018}. Both the \etac and the \jpsi mesons are reconstructed in the \ppbar final state. The measured ratio is determined as 
\begin{equation}
\begin{aligned}
\frac{\sigma^{\text{prompt (\bquark)}}_{\etac}}{\sigma^{\text{prompt (\bquark)}}_{\jpsi}} &= \frac{N^{\text{prompt (\bquark)}}_{\etac}}{N^{\text{prompt (\bquark)}}_{\jpsi}}\times \frac{\epsilon_{\jpsi}}{\epsilon_{\etac}}\times \frac{\BR_{\JpsiToPpbar}}{\BR_{\EtacToPpbar}},
\end{aligned}
\label{eq:eq1}
\end{equation}
where $\sigma^{\text{prompt}}_{\etac}$ is the $\etac$ prompt production cross-section and $\sigma^{\bquark}_{\etac}$ is the production cross-section in inclusive \bquark-hadron decays; 
$N^{\text{prompt}}_{\etac}$ and $N^{\bquark}_{\etac}$ are the signal yields of \etac mesons produced promptly and in \bquark-hadron decays, respectively. 
Similar definitions apply for the \jpsi yields and cross-sections.
The $\frac{\epsilon_{\jpsi}}{\epsilon_{\etac}}$ is the ratio of total efficiencies to trigger, reconstruct and select \JpsiToPpbar and \EtacToPpbar decays, which is found to be the same, within uncertainties, for prompt and \bquark-decay charmonia. 
The ratio of branching fractions of \bquark-hadron inclusive decays to \etac and to \jpsi mesons,  $\frac{\BR_{\bToEtacX}}{\BR_{\bToJpsiX}}=\frac{\sigma^{\bquark}_{\etac}}{\sigma^{\bquark}_{\jpsi}}$, is defined in the same way as for prompt production.
The values of the branching fractions of the 
\etac and \jpsi decays to \ppbar,  
$\BR_{\EtacToPpbar}=\brEtacpp$ and $\BR_{\JpsiToPpbar}=\brJpsipp$, correspond to the current world-average values~\cite{PDG2018}.
While the branching fraction of \bquark-hadron inclusive decays to \jpsi meson, 
$\BR_{\bToJpsiX} = (1.16\pm0.10) \times 10^{-3}$, was measured at \lep~\cite{PDG2018}, this analysis assumes the same value for the \bquark-hadron mixture at \lhc.

Since the masses of the \etac and \jpsi states and kinematic distributions in \JpsiToPpbar and \EtacToPpbar decays are similar, they have similar reconstruction, trigger and selection efficiencies.
Using simulation, the efficiency ratio of the \JpsiToPpbar and \EtacToPpbar modes is determined  to be $\epsilon_{\jpsi}/\epsilon_{\etac} = 1.00\pm0.02$,
where the uncertainty is due to the size of the simulation samples. The efficiency ratio is also obtained in bins of \pt, with negligible 
deviation from unity observed. Prompt \jpsi mesons are assumed to be unpolarised. A systematic uncertainty is further assigned related to possible non-zero polarisation.

In the baseline analysis, promptly produced charmonium candidates are 
distinguished from those originating from \bquark-hadron decays using the pseudo-proper decay time
 \begin{equation}
    t_z = \frac{\Delta z \cdot M_{\ppbar}}{p_{z}}\, ,
 \end{equation}
where $\Delta z$ is the distance along the beam axis between the PV with the smallest IP significance of the charmonium candidate and the charmonium decay vertex; 
$M_{\ppbar}$ is the reconstructed charmonium mass; 
and ${p_{z}}$ is the projection of its momentum along the beam axis.
A prompt-enriched sample is selected with candidates satisfying $t_z < 80 \fs$, while a \bquark-hadron-enriched sample is selected with $t_z > 80 \fs$ and an additional requirement that both proton tracks are significantly displaced from any PV.
These two samples have a small percentage of wrongly classified candidates. The probability of such cross-feed is estimated using simulation and is used to derive corrected yields. The cross-feed correction on the yield ratio ranges from 1.1\% to 2.7\% in the prompt-enriched sample and 0.7\% to 1.2\% in the \bquark-hadron-enriched sample, depending on the charmonium \pt.

A cross-check of the results, reported in Appendix~\ref{sec:tzfit},  uses an alternative approach analysing the $t_z$ distribution of the selected candidates. 
The results are in good agreement with the baseline analysis.

\section{Fit to the invariant mass}
\label{sec:massFit}
A binned fit to the \ppbar invariant mass of the prompt-enriched and \bquark-hadron-enriched data samples is performed simultaneously in each bin of charmonium \pt in order to extract the \jpsi signal yield and the \etac-to-\jpsi yield ratio.
For the \etac state, the signal shape is modelled by a relativistic Breit--Wigner function convolved with the sum of two Gaussian functions, while the signal shape of the \jpsi state is modelled by the sum of two Gaussian functions. In the study of the \etac production, the mass values 
$M_{\jpsi}$ and $\Delta M_{\jpsi,\,\etac}$ 
are constrained within uncertainties in each \pt bin to the values obtained from a fit to the entire data sample, where they are found to be consistent with the known values~\cite{PDG2018}.
The mass resolutions of charmonium from \bquark-hadron decays and from prompt production are assumed to be the same, as confirmed by simulation.
The ratio between the widths of the resolution functions for the \jpsi and \etac mesons is fixed from simulation. The only resolution parameter left free to vary in the fit 
is the width of the narrower Gaussian in the  \etac model. 
A small \pt-dependence of resolution parameters is seen in the simulation and is accounted for in the fit. 
The natural width of the \etac meson is fixed to its known value~\cite{PDG2018}. 
The combinatorial background is parametrised using an exponential multiplied by a second-order polynomial.
A partially reconstructed background of proton-antiproton pairs from the decays of higher mass charmonium states 
could exhibit structures in the $\ppbar$ invariant mass spectrum. The only contribution 
relevant for this analysis is that from $\JpsiToPpbarPiz$ decays, where the \piz meson is not reconstructed. 
This background produces a broad, non peaking, contribution to the \ppbar invariant mass below the known \etac mass. In this region, the $\ppbar\piz$ background is described by a square-root shape, $\sqrt{M_T-M_{\ppbar}}$, below the phase-space limit, $M_T$.
The yield of this contribution is related to that of the decay ${\jpsi}\to{\ppbar}$ by means of the
ratio of branching fractions $\BR_{\JpsiToPpbarPiz} / \BR_{\JpsiToPpbar} = \brRatioJpsipppizJpsipp$~\cite{PDG2018} and the ratio of efficiencies $\epsilon_{\JpsiToPpbarPiz}/\epsilon_{\JpsiToPpbar} = \epsRatioJpsipppizJpsipp$ for considered \ppbar invariant mass window.

\begin{figure}[t]
\centering
\protect\includegraphics[width=1.0\textwidth]{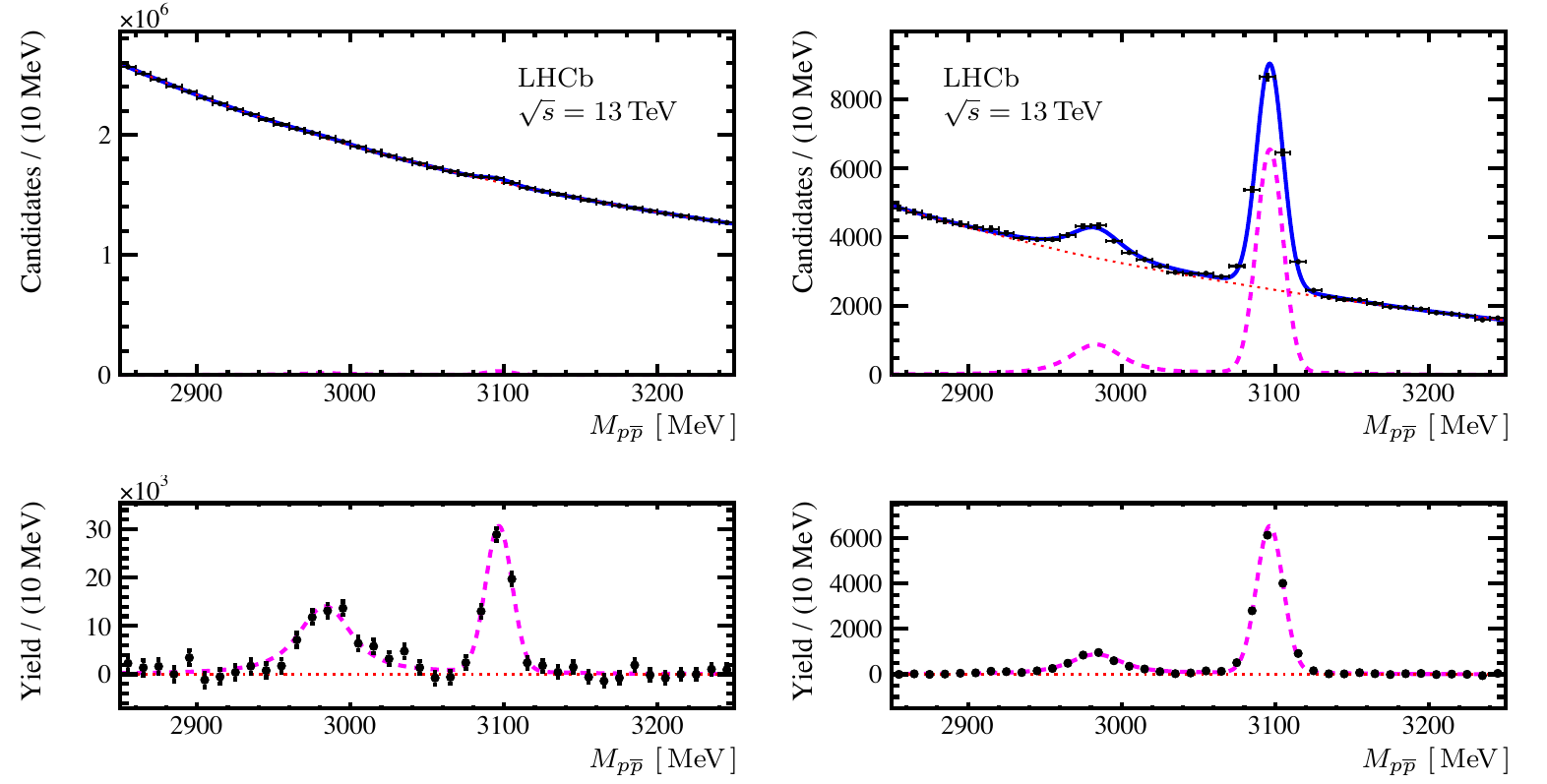}
\caption{Invariant-mass distribution of the \ppbar candidates for (top left) prompt-enriched  and (top right) \bquark-hadron-enriched samples in the entire \pt range, $6.5<\pt<14.0 \gev$. The solid blue lines represent the total fit result. The dashed magenta and dotted red lines show the signal and background components, respectively. Signal distributions with the background components from the fit subtracted are shown on the bottom plots for the two samples.} 
\label{fig:massFitRunIInt}
\end{figure}

The \ppbar invariant mass of selected candidates is shown in Fig.~\ref{fig:massFitRunIInt}. 
Projections of the simultaneous fit result integrated over the entire \pt range are overlaid. 
In general, the fit provides a good description of all $M_{\ppbar}$ distributions. The charmonium yields in bins of \pt and for the entire data sample are summarised in Table~\ref{tab:yieldsTzCut}. These yields are corrected to take into account the cross-feed probabilities.

\begin{table}[b]
\caption{Yield of \jpsi mesons and the 
\etac-to-\jpsi yield ratio for prompt and \bquark-hadron decay production, corrected for the cross-feed, in bins of transverse momentum.}
\label{tab:yieldsTzCut}
\centering
\small
\begin{tabular}{l|c|c|c|c} 
\pt range $[\gev\,]$ & $N^{\text{prompt}}_{\jpsi}$ & $N^{\bquark}_{\jpsi}$ & $\frac{N^{\text{prompt}}_{\etac}}{N^{\text{prompt}}_{\jpsi}}$ & $\frac{N^{\bquark}_{\etac}}{N^{\bquark}_{\jpsi}}$  \\ \hline
\phantom{1}$6.5-\phantom{1}8.0$  & $21600\pm1800$ & \phantom{1}$5080\pm140$  & $0.98\pm0.22$ & $0.26\pm0.04$ \\
\phantom{1}$8.0-10.0$ & $26500\pm1700$ & \phantom{1}$7930\pm170$  & $1.12\pm0.18$ & $0.40\pm0.03$ \\
$10.0-12.0$           & $15100\pm1100$ & \phantom{1}$5240\pm130$  & $1.24\pm0.19$ & $0.30\pm0.04$ \\
$12.0-14.0$           & $\phantom{1}5700\pm\phantom{1}700$   & $\phantom{1}2830\pm100$   & $2.24\pm0.44$ & $0.35\pm0.05$ \\ \hline
\phantom{1}$6.5-14.0$ & $69000\pm2800$ & $21040\pm270$ & $1.18\pm0.10$ & $0.33\pm0.02$ \\
\hline \hline
\end{tabular} 
\end{table}

\section{Systematic uncertainties} 
\label{sec:syst}
The systematic uncertainties on the \etac production corresponding to the  signal and background descriptions in the invariant-mass fit are estimated using alternative fit models. Each uncertainty is estimated in \pt bins as a difference between the baseline fit result and the alternative fit result. 
Generally, bin-to-bin variations of uncorrelated systematic uncertainties addressed in this paragraph are small compared to the statistical uncertainties.
These variations are parametrised using a linear \pt dependence to reduce statistical fluctuations.
The uncertainty related to a potential \pt dependence of 
the resolution ratio
is evaluated by modelling it using a linear function with the slope constrained by simulation. 
This systematic effect is relevant for the differential cross-section measurement.
The uncertainty corresponding to the combinatorial background description is estimated by using an alternative model using a third-order polynomial function. 
The uncertainty associated to the \JpsiToPpbarPiz background contribution is estimated by varying the efficiency ratio $\epsilon_{\JpsiToPpbarPiz}/\epsilon_{\JpsiToPpbar}$
and the branching fraction ratio $\BR_{\JpsiToPpbar}/\BR_{\JpsiToPpbarPiz}$~\cite{PDG2018} within their uncertainties. 

The systematic uncertainty associated with the cross-feed probability is estimated by varying the efficiency of the separation requirements. 
Separation efficiencies are found to be in good agreement between data and simulation within uncertainties.
The uncertainty related to the ratio of \etac and \jpsi efficiencies is estimated by varying its value by the uncertainty corresponding to the simulation sample sizes.

A potential effect due to the invariant-mass resolution modelling is evaluated considering, as an alternative model, the sum of two Crystal Ball functions~\cite{Skwarnicki:1986xj} with symmetric tails on both sides of the peak.
The uncertainty on the \etac natural width is accounted for by the difference in relative yields 
when using the world average value of $31.9 \pm 0.7$\mev~\cite{PDG2018} and the value, $34.0 \pm 1.9 \pm 1.3$\mev, recently measured by \lhcb collaboration~\cite{LHCb-PAPER-2016-016}. 
Since this uncertainty is correlated among \pt bins, the relative systematic uncertainty obtained from the \pt-integrated data sample is taken as an estimate of the relative systematic uncertainty in each bin.
This uncertainty is also correlated between \pt bins.
Possible nonzero polarisation of prompt \jpsi mesons affects their reconstruction efficiency. The \jpsi polarisation has not been measured at \sqs=13\tev, although several experiments have measured small polarisation values at lower energy~\cite{LHCb-PAPER-2013-008,Abelev:2011md,Chatrchyan:2013cla}. The associated systematic uncertainty is estimated by weighting the prompt \jpsi simulation sample assuming polarisation parameter values $\lambda_{\theta}=\pm0.1$~\cite{Jacob:1959at} in the \proton\proton collision frame. This uncertainty is correlated among \pt bins.

Systematic uncertainties on the 
relative cross-sections of the \etac production
for prompt and \bquark-hadron decays are given in Tables~\ref{tab:systPTAll} and~\ref{tab:systRunITotal}. 
The total systematic uncertainty is calculated as the quadratic sum of the individual sources.
The dominant source of uncorrelated systematic uncertainty 
for the production of \etac meson for both prompt and from \bquark-hadron decays is related to the combinatorial background description.
The dominant sources of correlated systematic uncertainty for prompt production are related to the knowledge of the \etac natural width and the invariant-mass resolution model. 
The uncertainty on the
knowledge of the \etac natural width is the dominant source of correlated systematic uncertainty for
\bquark-hadron decay production.
Systematic uncertainties are in general smaller than the corresponding statistical uncertainties.
\begin{table}[t]
\centering
\small
\caption{Relative uncertainties (in \%) on the ratio of prompt cross-sections $\sigma^{\text{prompt}}_{\etac}/\sigma^{\text{prompt}}_{\jpsi}$. Uncertainties on $\BR_{\EtacToPpbar}$ and  $\BR_{\JpsiToPpbar}$ are considered separately and given in the text.}
\label{tab:systPTAll}
\addtolength{\tabcolsep}{-1pt} 
\begin{tabular}{l|c|c|c|c||c} 
                         & \multicolumn{5}{c}{\pt range $[\gev\,]$}\\ \cline{2-6}
                         & $6.5-8.0$ & $8.0-10.0$ & $10.0-12.0$ & $12.0-14.0$ & $6.5-14.0$ \\ \hline 
Stat. unc.                                  & \!\!22.7  & \!\!16.1  & \!\!16.9  & \!\!18.3 & 8.8     \\ \hline \hline
\pt dependence of resolution       & 0.4   & 0.4   & 0.4   & 0.4 & 0.2     \\  
Comb. bkg. description        & 2.1   & 3.3   & 4.6   & 6.0 & 2.0      \\ 
Contribution from \JpsiToPpbarPiz & 0.2   & 0.2   & 0.3   & 0.3 & $<0.1 \ \ \ $   \\    
Cross-feed                                             & 1.9   & 1.1   & 1.2   & 1.4 & 0.9      \\   
Efficiency ratio                                             & 2.0   & 2.0   & 2.0   & 2.0 & 2.0      \\   
\hline  
Total uncorrelated syst. unc.             & 3.5   & 4.0   & 5.2   & 6.5 & 3.0      \\   
\hline  
\hline  
Mass resolution model                      & 2.7   & 2.7   & 2.7   & 2.7 & 2.7      \\   
Variation of $\Gamma_{\etac}$                               		  & 4.8   & 4.8   & 4.8   & 4.8 & 4.8      \\ 
\jpsi polarisation                                & 2.1   & 1.8   & 1.6   & 1.6 & 1.8      \\    
\hline  
Total correlated syst. unc.                & 5.8   & 5.7   & 5.7   & 5.7 & 5.7      \\   
\hline  
\hline  
Total systematic uncertainty                                  & 6.8   & 7.0   & 7.7   & 8.6 & 6.4      \\ 
\hline  
\hline
\end{tabular} 
\addtolength{\tabcolsep}{1pt} 
\end{table} 
\begin{table}[h!]
\centering
\small
\caption{Relative uncertainties (in \%) on the ratio of cross-sections for production in inclusive \bquark-hadron decays $\sigma^{b}_{\etac}/\sigma^{b}_{\jpsi}$. Uncertainties on $\BR_{\EtacToPpbar}$ and  $\BR_{\JpsiToPpbar}$ are considered separately and given in the text.}
\label{tab:systRunITotal}
\addtolength{\tabcolsep}{-1pt}
\begin{tabular}{l|c|c|c|c||c} 
                         & \multicolumn{5}{c}{\pt range $[\gev\,]$}\\ \cline{2-6}
                         & $6.5-8.0$ & $8.0-10.0$ & $10.0-12.0$ & $12.0-14.0$ & $6.5-14.0$ \\ \hline 
Stat. unc.                                 & \!\!15.4  & 8.2   & 1\!\!2.8  & \!\!13.4 & 5.8     \\ \hline \hline
\pt dependence of resolution & 0.2   & 0.2   & 0.2   & 0.2 & 0.1      \\  
Comb. bkg. description                    & 2.5   & 3.5   & 4.7   & 5.8 & 2.3      \\ 
Contribution from \JpsiToPpbarPiz                 & 0.7   & 0.5   & 0.3   & 0.1 & 0.2      \\    
Cross-feed                                        & 1.4   & 1.3   & 1.7   & 1.0 & 0.8      \\   
Efficiency ratio                                             & 2.0   & 2.0   & 2.0   & 2.0 & 2.0      \\   
\hline  
Total uncorrelated syst. unc.                     & 3.6   & 4.3   & 5.4   & 6.2 & 3.2      \\   
\hline  
\hline  
Mass resolution model                             & 3.1   & 3.1   & 3.1   & 3.1 & 3.1      \\   
Variation of $\Gamma_{\etac}$                     & 3.6   & 3.6   & 3.6   & 3.6 & 3.6      \\ 
\hline  
Total correlated syst. unc.                      & 4.8   & 4.8   & 4.8   & 4.8 & 4.8      \\   
\hline  
\hline  
Total systematic uncertainty                                  & 6.0   & 6.4   & 7.2   & 7.8 & 5.8      \\
\hline  
\hline 
\end{tabular}
\addtolength{\tabcolsep}{1pt} 
\end{table} 

Uncertainties on the branching fractions of the \JpsiToPpbar 
and \EtacToPpbar  decays 
are considered separately. They are correlated among \pt bins and amount to about 10\%.
When deriving the absolute $\etac$ production cross-section, the uncertainty on the \jpsi production cross-section~\cite{LHCb-PAPER-2015-037} is also taken into account.

%% file: tzcut.tex
\section{Results and discussion on the \pmb{\etac} production}
\label{sec:results}
Using Eq.~\ref{eq:eq1} and the corrected yields from Table~\ref{tab:yieldsTzCut},
the relative prompt production cross-section in the chosen fiducial region is measured to be 
\begin{align*}
\left( \sigma_{\etac}^{\text{prompt}}/\sigma_{\jpsi}^{\text{prompt}} \right)_{13 \tev}^{6.5 < \pt < 14.0 \gev,\,2.0<y<4.5} 
  &= \etacPromptRelativeXsec.
\end{align*}
Here and hereinafter, the quoted uncertainties are statistical, systematic and systematic due to uncertainties on the branching fractions $\BR_{\JpsiToPpbar}$ and $\BR_{\EtacToPpbar}$, respectively.
Using the corresponding cross-section value of $\jpsiPromptAbsoluteXsec$ 
for prompt \jpsi production~\cite{LHCb-PAPER-2015-037}, 
the prompt \etac production cross-section in the chosen fiducial region is derived to be 
\begin{align*}
\left( \sigma_{\etac}^{\text{prompt}} \right)_{13 \tev}^{6.5 < \pt < 14.0 \gev,\,2.0<y<4.5}  
  &= \etacPromptAbsoluteXsec,
\end{align*}
where the last uncertainty includes in addition the uncertainty on the \jpsi production cross-section measurement. This is the first measurement of prompt \etac production cross-section in proton-proton collisions at \sqs=13\tev, and it supports the conclusions of Ref.~\cite{LHCb-PAPER-2014-029} that suggests an enhanced \etac production compared to that of the \jpsi meson.
This measurement  
is in good agreement with the colour-singlet model prediction of $1.56^{+0.83}_{-0.49}\,^{+0.38}_{-0.17}\mub$~\cite{Feng:2019zmn}. 
This result leaves limited room for a potential colour-octet contribution, and confirms the theoretical analyses~\cite{Han:2014jya,Bodwin:2014gia,Zhang:2014ybe,Butenschoen:2014dra} following the \etac production studies at ${\sqs=7}$ and ${8 \tev}$~\cite{LHCb-PAPER-2014-029}.

\begin{figure}[t]
\protect\includegraphics[width=1.0\textwidth]{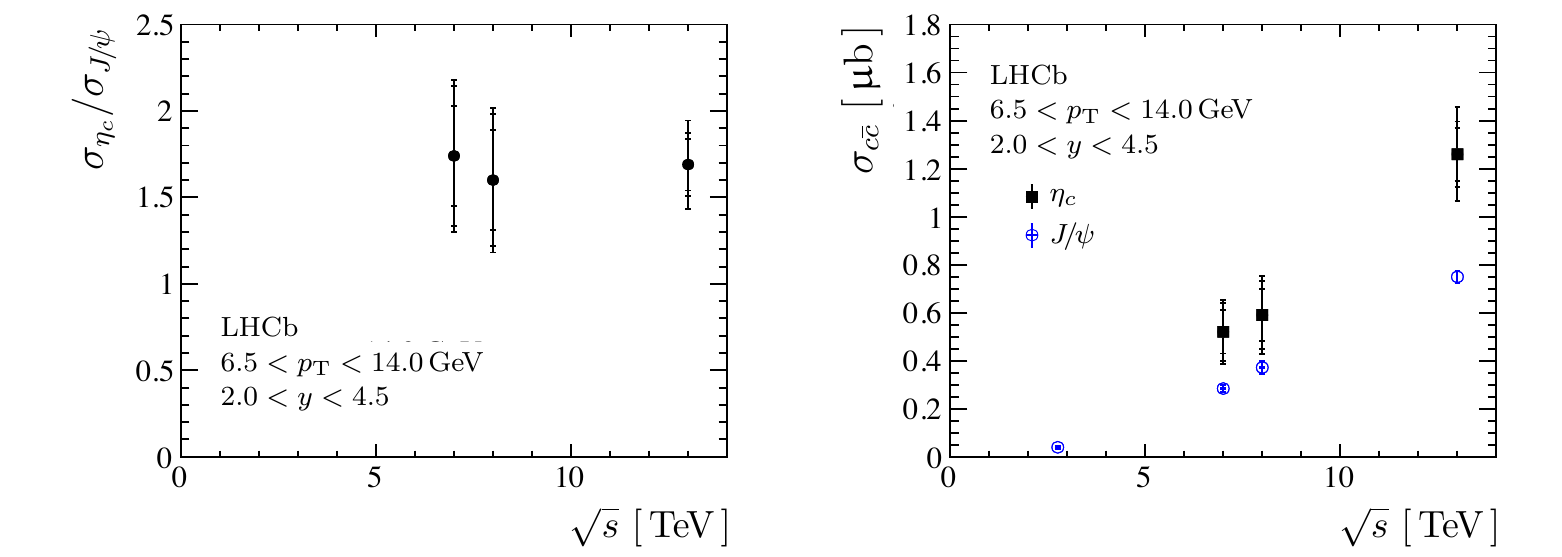}
\caption{The prompt \etac production cross-section as a function of centre-of-mass energy. (Left) Relative \etac prompt production cross-section. (Right) Absolute \etac (black rectangles) and \jpsi (blue circles) prompt production cross-sections. The error bars show uncertainties due to statistical, systematic, and to the \EtacToPpbar and \JpsiToPpbar branching fractions and \jpsi production cross-section.}
\label{fig:Xsec_vs_s}
\end{figure}

Using the \lhcb measurements of prompt \etac production at the centre-of-mass energies ${\sqs=7}$ and ${8 \tev}$~\cite{LHCb-PAPER-2014-029}, the prompt \etac production cross-section dependence on the \lhc energy is shown in Fig.~\ref{fig:Xsec_vs_s}.
The \jpsi production cross-section from Ref.~\cite{LHCb-PAPER-2015-037} is also shown for reference. While the individual cross-sections grow with centre-of-mass energy, no evolution of the cross-section ratio is observed.

The relative \etac inclusive branching fraction from \bquark-hadron decays is measured to be
\begin{align*}
\BR_{\bToEtacX}/\BR_{\bToJpsiX} &= \etacSecondaryRelativeBR, 
\end{align*}
which combined with $\BR_{\bToJpsiX}=1.16\pm0.10 \%$~\cite{PDG2018} gives
\begin{align*}
\BR_{\bToEtacX} &= \etacSecondaryAbsoluteBR. 
\end{align*}
The last uncertainty includes the uncertainty on $\BR_{\bToJpsiX}$. This result is the most precise measurement of the inclusive \bToEtacX branching fraction to date and is in good agreement with the previous \lhcb measurement from Ref.~\cite{LHCb-PAPER-2014-029}. The measurement is limited by the knowledge of the branching fractions $\BR_{\EtacToPpbar}$ and $\BR_{\bToJpsiX}$.

\begin{figure}[h]
\protect\includegraphics[width=1.0\textwidth]{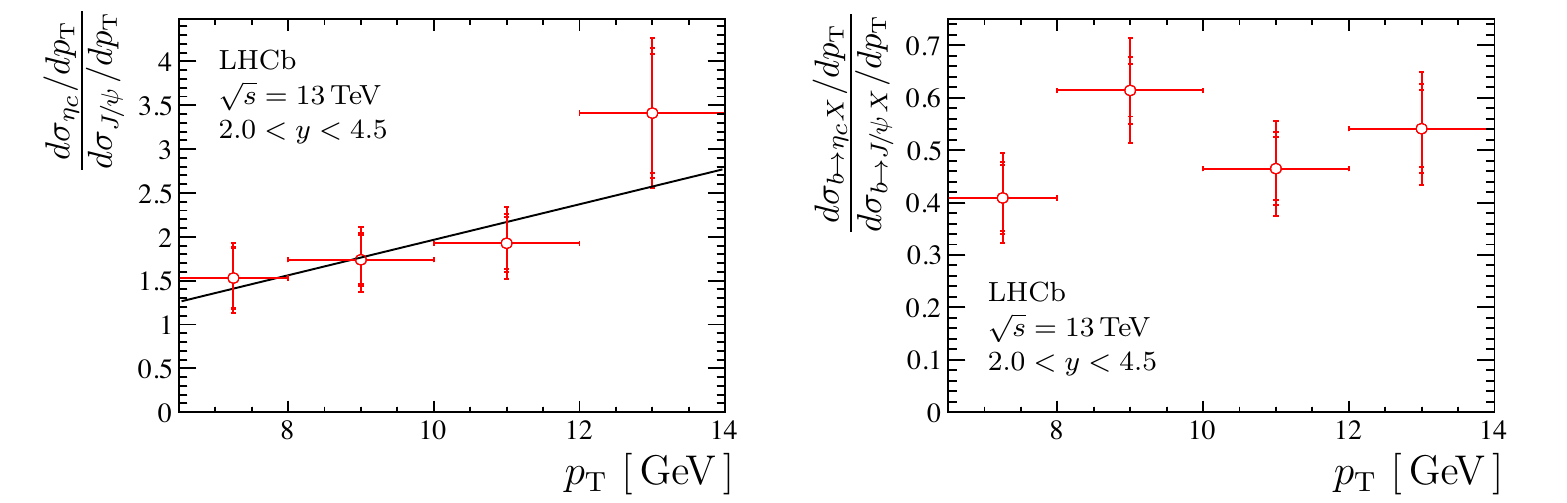}
\caption{Relative \etac to \jpsi differential production cross-sections for (left) prompt production and (right) production in \bquark-hadron inclusive decays. The uncertainties are statistical, systematic, and due to the \EtacToPpbar and \JpsiToPpbar branching fractions, respectively. For the relative prompt production cross-section, the result of a fit with a linear function is overlaid.} 
\label{fig:sigmaRel}
\end{figure}

\begin{figure}[t]
\protect\includegraphics[width=1.0\textwidth]{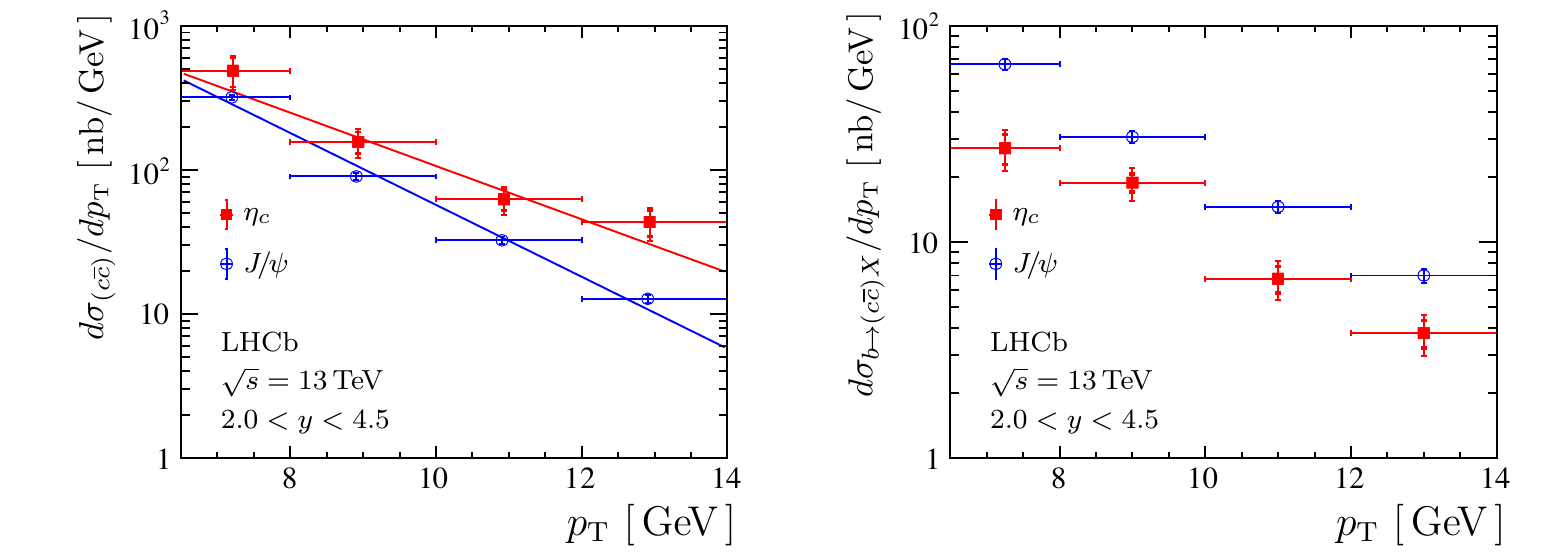}
\caption{Differential production cross-sections of \etac (red rectangles) and \jpsi (blue circles) for (left) prompt production and (right) production in  \bquark-hadron inclusive decays. The uncertainties for \etac production are statistical, systematic, and due to the \EtacToPpbar and \JpsiToPpbar branching fractions and \jpsi production cross-section. For the prompt production cross-sections, the results of fits with an exponential function are overlaid. The \pt values of the data points correspond to the average values of the fit function over the bins.} 
\label{fig:sigmaAbs}
\end{figure}

Numerical results of the measurements of \pt-differential \etac production are given in Appendix~\ref{app:sigmaRel}.
The relative \etac to \jpsi \pt-differential cross-sections for prompt and \bquark-hadron decay production are compatible to those measured at ${\sqs=7}$ and ${8 \tev}$~\cite{LHCb-PAPER-2014-029} and are shown in Fig.~\ref{fig:sigmaRel}.
This is the first \pt-differential cross-section measurement of \etac prompt production at ${\sqs=13 \tev}$. The \pt dependence of the prompt cross-section ratio is found to be linear with a slope of $0.22\pm0.11 \gev^{-1}$. While the integrated cross-section is in good agreement with the colour-singlet model prediction~\cite{Feng:2019zmn}, a hint of a difference between the \jpsi and \etac slopes motivates the extension of the measurement to larger \pt values.
A larger measured slope with respect to the prediction from Ref.~\cite{Feng:2019zmn} would indicate a possible colour-octet contribution.
The absolute \etac and \jpsi differential production cross-sections are shown in Fig.~\ref{fig:sigmaAbs}. The exponential slopes for the \etac and \jpsi prompt differential cross-sections are determined from the fit to data points to be $0.41\pm0.07 \gev^{-1}$ and $0.57\pm0.01 \gev^{-1}$, respectively. 

%% file: mass.tex
\section{Measurement of the $\pmb{\jpsi}$--$\pmb{\etac}$ mass difference}
\label{sec:mass}
While the prompt $\etac$ production measurement requires 
stringent selection criteria at the trigger level to compete with the challenging background conditions,
charmonia produced in \bquark-hadron decays are reconstructed in an environment with a controlled background level and are more suitable for a mass measurement. 
For this reason, a looser selection is applied for the entire data sample to measure the $\etac$ mass relative to the well-known \jpsi mass.

Proton and antiproton candidates are required to have good track-fit quality,
to be incompatible with originating from any PV,
and to have a transverse momentum greater than 1.0\gev. 
The proton-antiproton system is required to have a vertex with a good fit quality, a large significance, $\chisq_{\text{FD}}>81$, of the distance between this vertex and any PV, and to have a transverse momentum greater than $5.5 \gev$.
The contamination of the selected sample from \jpsi and \etac prompt production is estimated to be below $0.1 \%$. 

\begin{figure}[t]
\protect\includegraphics[width=1.0\textwidth]{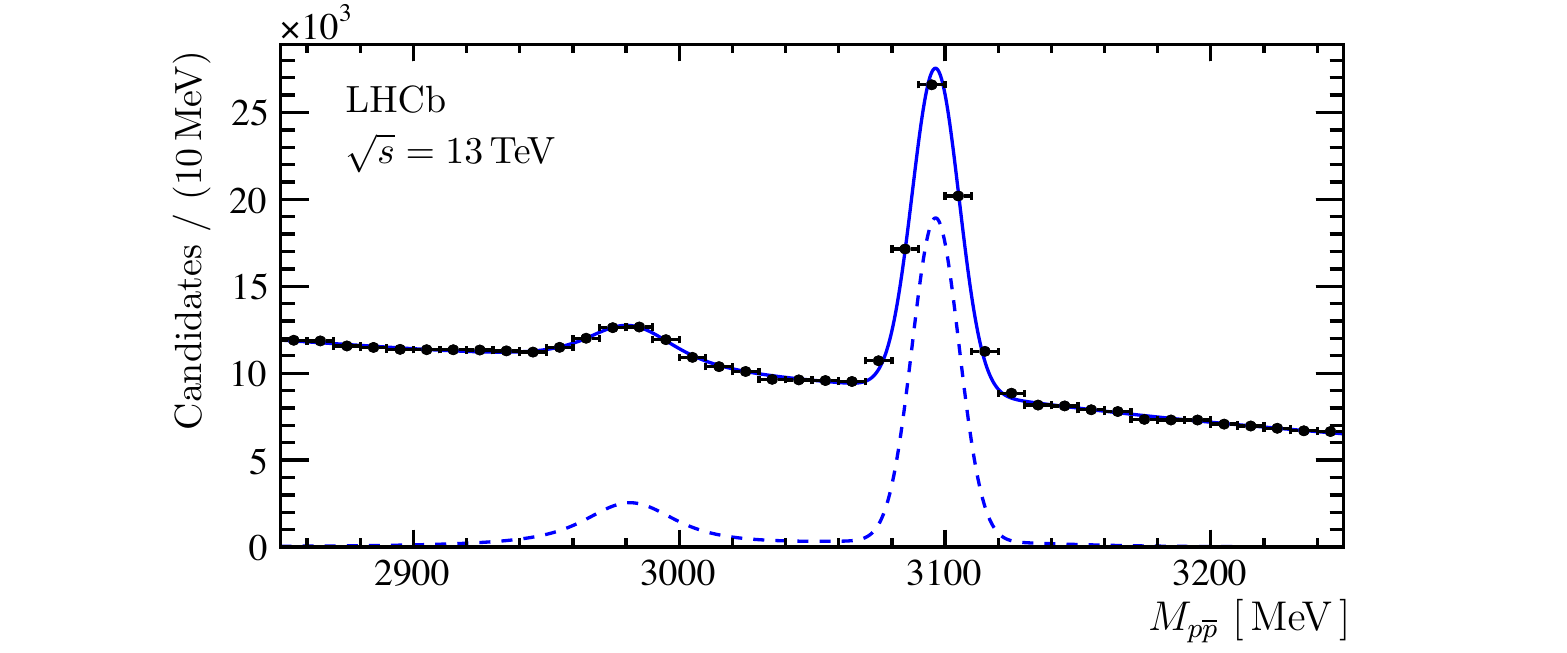}
\caption{Distribution of the $\ppbar$ invariant mass. The solid line represents the fit result, while the dashed line shows the \etac and \jpsi signal contributions.} 
\label{fig:massTopo}
\end{figure}

The mass difference $\Delta M _{\jpsi , \, \etac}$ is extracted from an extended maximum-likelihood fit to the $M_{\ppbar}$ distribution. 
The fit provides a good description of the \ppbar invariant-mass distribution (Fig.~\ref{fig:massTopo}) yielding
\begin{equation*}
\Delta M _{\jpsi , \, \etac} = 113.0\pm0.7\pm0.1 \mev,
\end{equation*}
where the uncertainties are statistical and systematic.

The majority of the sources of systematic uncertainty are common to the production measurement. 
The systematic uncertainty related to the momentum-scale calibration is estimated by comparing the fit result with and without the calibration applied.
The total systematic uncertainty is calculated as the quadratic sum of the individual contributions (Table~\ref{tab:deltam}). The dominant source of systematic uncertainty is related to the resolution model and its \pt dependence. 
\begin{table}[t]
\caption{Systematic uncertainties on the measurement of the \jpsi--$\etac$ mass difference.} 
\label{tab:deltam}
\begin{center} 
\begin{tabular}{l|*{1}{r}} 
 			                 				& $\Delta M _{\jpsi , \, \etac}\,[\mev\,]$ \\ \hline 
Statistical uncertainty	     				    & $0.67$\phantom{00000} 		    	\\ \hline \hline
Mass resolution model  	                    & $0.08$\phantom{00000}	 		    \\ 
Variation of resolution ratio& $0.01$\phantom{00000}        \\
Variation of $\Gamma_\etac$                & $0.04$\phantom{00000}				\\
Comb. bkg. description                      & $0.03$\phantom{00000}			  \\
Contribution from \JpsiToPpbarPiz           & $<0.01$\phantom{00000}      \\  
Momentum scale                              & $0.05$\phantom{00000}    \\ \hline \hline
Total systematic uncertainty                & $0.11$\phantom{00000}	 	\\ \hline \hline
Total uncertainty 							& $0.68$\phantom{00000}  	\\ \hline \hline
\end{tabular} 
\end{center} 
\end{table} 

As a cross-check, the invariant-mass fit is performed simultaneously in seven bins of charmonium transverse momentum to take into account a possible dependence of the resolution on charmonium \pt. The value obtained for the mass difference is consistent with the baseline result.

This measurement is currently statistically limited and can be improved with larger data samples.
It represents the most precise measurement from a single experiment to date. The result is in good agreement with the PDG value~\cite{PDG2018}, the recent BES III result~\cite{BESIII:2011ab}, the latest \babar measurement~\cite{Lees:2014iua} and \lhcb measurements~\cite{LHCb-PAPER-2014-029, LHCb-PAPER-2017-007, LHCb-PAPER-2016-016}.  

%% file: summary.tex
\section{Summary}
\label{sec:summary}
Using data corresponding to an integrated luminosity of 2.0\invfb, 
the prompt $\etac$ production cross-section at a centre-of-mass energy of $\sqs=13$\tev is measured for the first time. 
The ratio of the prompt production rates of the $\etac$ and \jpsi states in the fiducial region ${6.5 < \pt < 14.0 \gev}$ and ${2.0<y<4.5}$ is measured to be 
\begin{align*}
\left( \sigma_{\etac}^{\text{prompt}}/\sigma_{\jpsi}^{\text{prompt}} \right)_{13 \tev}^{6.5 < \pt < 14.0 \gev,\,2.0<y<4.5}  
    &= \etacPromptRelativeXsec, 
\end{align*}
where the quoted uncertainties are, in order, statistical, systematic and systematic due to uncertainties on the branching fractions, $\BR_{\JpsiToPpbar}$ and $\BR_{\EtacToPpbar}$.

Using the prompt \jpsi production cross-section measurement at $\sqs=13 \tev$~\cite{LHCb-PAPER-2015-037}, the prompt $\etac$ production cross-section in the chosen fiducial region is derived to be 
\begin{align*}
\left( \sigma_{\etac}^{\text{prompt}} \right)_{13 \tev}^{6.5 < \pt < 14.0 \gev,\,2.0<y<4.5}  
    &= \etacPromptAbsoluteXsec,
\end{align*} 
where the last uncertainty includes in addition the uncertainty of the \jpsi production cross-section measurement.
The result is in good agreement with the colour-singlet model prediction~\cite{Feng:2019zmn}. Contrary to NRQCD expectations, a steeper \pt dependence of the \jpsi cross-section compared to that of the \etac is preferred.

The relative $\etac$ inclusive branching fraction from \bquark-hadron decays is measured to be
\begin{align*}
\BR_{\bToEtacX}/\BR_{\bToJpsiX} &= \etacSecondaryRelativeBR. 
\end{align*}
Using $\BR_{\bToJpsiX}$~\cite{PDG2018} the absolute \etac inclusive branching fraction is obtained to be
\begin{align*}
\BR_{\bToEtacX} &= \etacSecondaryAbsoluteBR, 
\end{align*}
where the last uncertainty includes in addition the uncertainty on $\BR_{\bToJpsiX}$.
This result is consistent with the previous \lhcb measurement~\cite{LHCb-PAPER-2014-029}.
Compatible results are obtained with an alternative analysis technique.

The \jpsi--$\etac$ mass difference is measured using an enlarged data sample of \bToEtacX decays. 
The measurement, 
\begin{align*}
\Delta M_{\jpsi , \, \etac} = \etacMassDiff, 
\end{align*}
is compatible with both the result of Refs.~\cite{LHCb-PAPER-2016-016} and~\cite{PDG2018}.
It is the most precise $\etac$ mass determination to date.

%% file: acknowledgements.tex
\section*{Acknowledgements}
%
%
\noindent We thank J.-P. Lansberg and H.-S. Shao for useful discussions. We express our gratitude to our colleagues in the CERN
accelerator departments for the excellent performance of the LHC. We
thank the technical and administrative staff at the LHCb
institutes.
We acknowledge support from CERN and from the national agencies:
CAPES, CNPq, FAPERJ and FINEP (Brazil); 
MOST and NSFC (China); 
CNRS/IN2P3 (France); 
BMBF, DFG and MPG (Germany); 
INFN (Italy); 
NWO (Netherlands); 
MNiSW and NCN (Poland); 
MEN/IFA (Romania); 
MSHE (Russia); 
MinECo (Spain); 
SNSF and SER (Switzerland); 
NASU (Ukraine); 
STFC (United Kingdom); 
DOE NP and NSF (USA).
We acknowledge the computing resources that are provided by CERN, IN2P3
(France), KIT and DESY (Germany), INFN (Italy), SURF (Netherlands),
PIC (Spain), GridPP (United Kingdom), RRCKI and Yandex
LLC (Russia), CSCS (Switzerland), IFIN-HH (Romania), CBPF (Brazil),
PL-GRID (Poland) and OSC (USA).
We are indebted to the communities behind the multiple open-source
software packages on which we depend.
Individual groups or members have received support from
AvH Foundation (Germany);
EPLANET, Marie Sk\l{}odowska-Curie Actions and ERC (European Union);
ANR, Labex P2IO and OCEVU, and R\'{e}gion Auvergne-Rh\^{o}ne-Alpes (France);
Key Research Program of Frontier Sciences of CAS, CAS PIFI, and the Thousand Talents Program (China);
RFBR, RSF and Yandex LLC (Russia);
GVA, XuntaGal and GENCAT (Spain);
the Royal Society
and the Leverhulme Trust (United Kingdom).

%% file: appendix.tex

\clearpage
\section*{Appendices}

\appendix

\section{Alternative analysis based on \pmb{$t_z$} distributions}
\label{sec:tzfit}
A cross-check of the results is performed using an alternative approach via a two-step procedure. Signal yields in bins of \pt and $t_z$ are obtained from a simultaneous fit to the corresponding \ppbar invariant mass distributions of candidates in the bin.
Prompt and \bquark-decay charmonium contributions are then determined using a simultaneous \chisq fit to the resulting $t_z$ distributions in  \pt bins, similarly to Ref.~\cite{LHCb-PAPER-2015-037}.

\begin{sidewaysfigure}[ht]
\protect\includegraphics[width=1.0\textwidth]{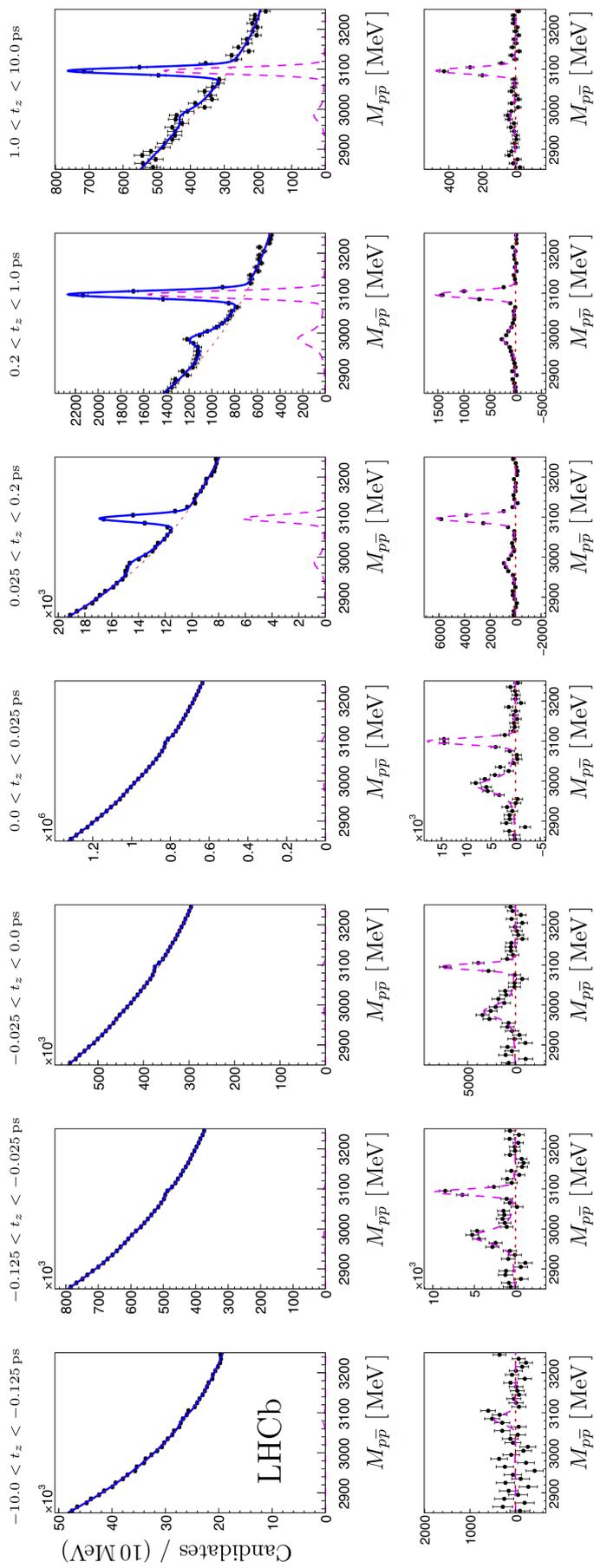}
\caption{Invariant mass distributions of the \ppbar candidates for seven bins of $t_z$ for the \pt-integrated sample $6.5<\pt<14.0 \gev$. The solid blue lines represent the total fit result. Dashed magenta and dotted red lines show the signal and background components, respectively. The distributions with the background component of the fit subtracted are shown below.} 
\label{fig:massFitInt}
\end{sidewaysfigure}

Projections of the simultaneous fit for the entire \pt-range are shown in Fig.~\ref{fig:massFitInt} for illustration purposes. 
In the alternative analysis discussed in this appendix, the \ppbar invariant-mass fit is performed simultaneously in 28 two-dimensional bins of \pt and $t_z$ with the model described in Sect.~\ref{sec:massFit}. 
This model is modified to correct for systematic mass shifts as a function of $t_z$. The corrections are derived from simulation, where the same behaviour is observed.\footnote{This effect cancels in the method described in Sect.~\ref{sec:massFit}, which integrates over all values of $t_z$.} 

\begin{figure}[t]
\protect\includegraphics[width=1.0\textwidth]{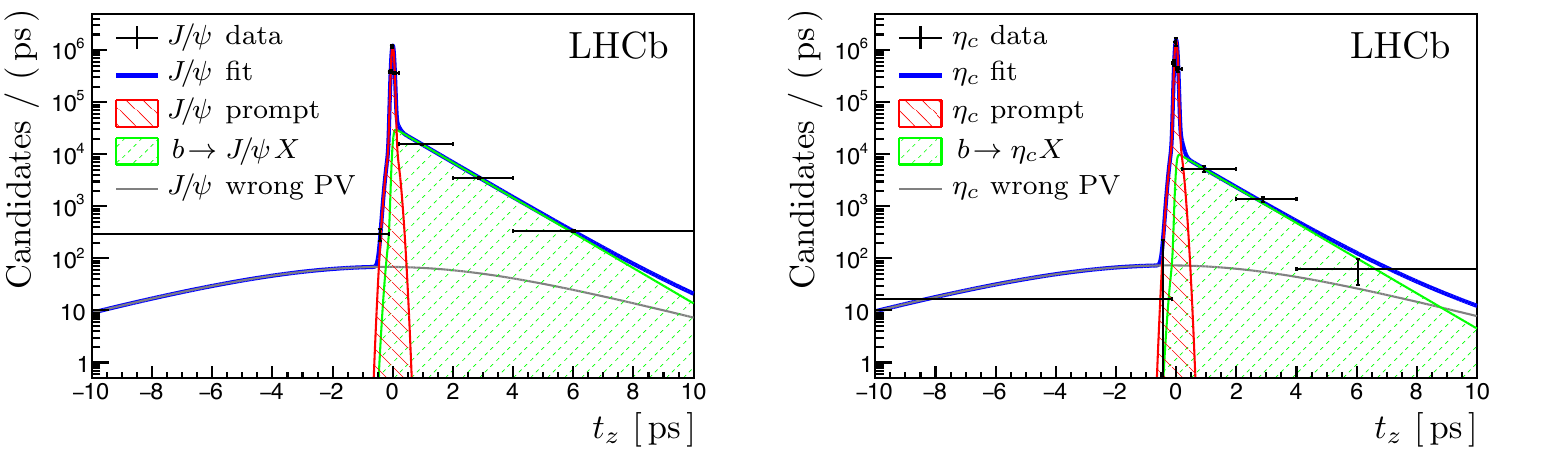}
\caption{Distributions of $t_z$ for (left) \jpsi and (right) $\etac$ mesons for the entire \pt range ${6.5<\pt<14.0 \gev}$ and the result of the simultaneous \chisq fit. 
The solid blue lines show the fit function, the red dashed areas show the prompt components, the green dotted areas show the contributions from inclusive \bquark-hadron decays, the grey lines show contributions from the events with wrongly associated PV. 
The $t_z$ values of the data points correspond to the average values of the fit function over the bins.} 
\label{fig:tzFitInt}
\end{figure}

The mass fits result in four $t_z$ distributions, each corresponding to a \pt bin. Promptly produced charmonium is distinguished from that produced in \bquark-hadron decays by performing a simultaneous \chisq fit to the four $t_z$ distributions.
This fit method does not use the bin centre for the value of $t_z$, but rather the average value of the fit function in the bin.
The model to describe the $t_z$ distribution comprises contributions due to prompt charmonia, due to charmonia from \bquark-hadron decays and a contribution due to candidates with a wrongly associated PV.
The prompt charmonium component is parametrised with a function to account only for 
resolution, while the component related to charmonia produced in inclusive \bquark-hadron decays is parametrised by an exponential decay function convolved with the same resolution function.
The exponential slope of the decay function, $\tau_b$, is allowed to vary over \pt according to simulation. 
The resolution is described by the sum of two Gaussian functions, 
with the width of the narrow Gaussian component a free fit parameter and the other parameters fixed from the simulation.
The \pt dependence of the resolution, as obtained from simulation, is taken into account in the $t_z$-fit to data.
The contribution due to candidates associated to a wrong PV are described in the same way as in Ref.~\cite{LHCb-PAPER-2015-037}.
Results of the simultaneous fit to $t_z$ for the entire \pt range are shown in Fig.~\ref{fig:tzFitInt}. 

In addition to the sources of systematic uncertainty discussed in Sect.~\ref{sec:syst} for the baseline analysis, contributions from the signal description in the fit to $t_z$ and
from corrections of the invariant-mass peak positions in $t_z$ bins are considered.
This approach is free from the systematic uncertainty associated with the cross-feed effect. The dominant sources of systematic uncertainties are the same as for the baseline analysis.

\begin{figure}[t]
\protect\includegraphics[width=1.0\textwidth]{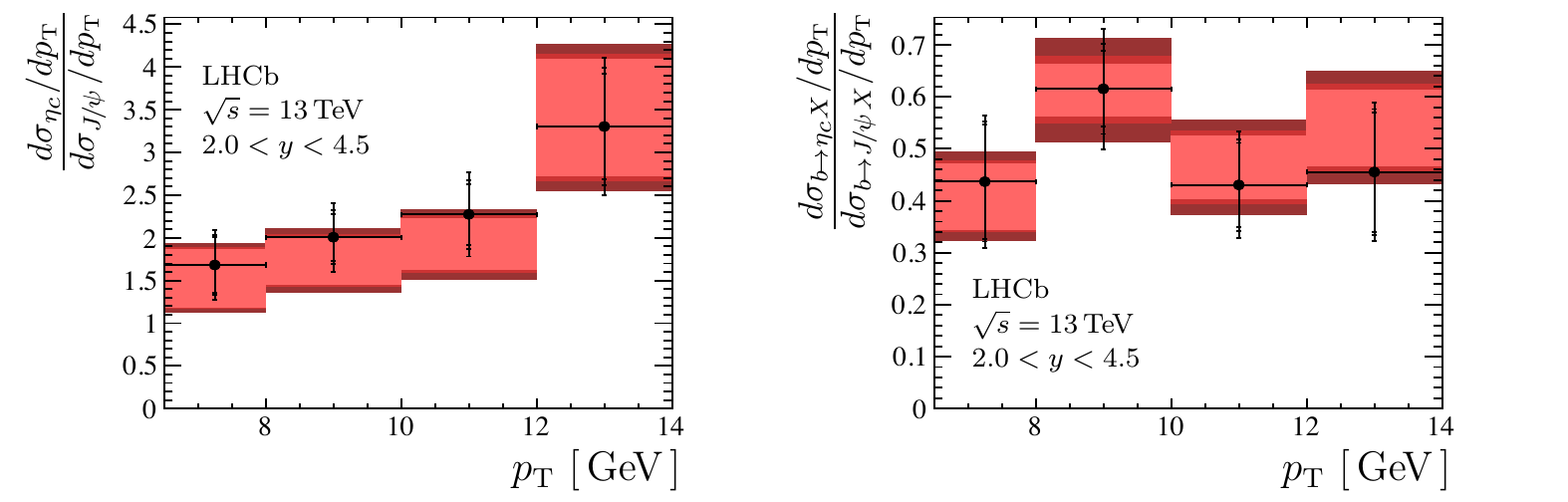}
\caption{Relative \etac to \jpsi differential production cross-sections for (left) prompt production and (right) production in \bquark-hadron inclusive decays obtained using the alternative technique (points) and the baseline technique (red boxes). The uncertainties shown are statistical, systematic, and the uncertainty due to the \EtacToPpbar and \JpsiToPpbar branching fractions.}
\label{fig:vsRunI}
\end{figure}

The relative differential cross-sections of \etac production obtained with
this approach are shown in Fig.~\ref{fig:vsRunI},  where they are compared
with those from the  baseline approach.
The two measurements are strongly correlated. The difference between the results obtained with the two techniques varies in \pt bins between factors of 0.5 and 1.5 of the estimated uncorrelated uncertainty. 

\clearpage
\section{Tables of \pmb{\pt}-differential \pmb{\etac} production cross-sections}
\label{app:sigmaRel}

\subsection{Prompt production of \pmb{\etac} mesons}
The results on relative \pt-differential \etac prompt production are shown in Table~\ref{tab:relPromptTable}. 
Here and in the following tables, the first uncertainty is statistical, the second is the uncorrelated systematic uncertainty, the third is the systematic uncertainty that is correlated among $\pt$ bins, and the last one is related to the 
 $\BR_{\JpsiToPpbar}$ and $\BR_{\EtacToPpbar}$ branching fractions.
The results on \pt-differential \etac prompt production cross-section 
are shown in Table~\ref{tab:PromptTable}. Here, the last uncertainty includes the uncertainty on the \jpsi production cross-section.
\begin{table}[h]
\centering
\caption{Relative \pt-differential \etac prompt production cross-section.}
\label{tab:relPromptTable}
\small
\begin{tabular}{c|c}
\centering
\pt $[\gev\,]$     & $d\sigma^{\text{prompt}}_{\etac}/d\sigma^{\text{prompt}}_{\jpsi}$ \\ \hline
$\phantom{1}6.5-\phantom{1}8.0$     & 1.53 $\pm$ 0.35 $\pm$ 0.05 $\pm$ 0.09 $\pm$ 0.19 \\ 
\phantom{1}$8.0-10.0$    & 1.74 $\pm$ 0.28 $\pm$ 0.07 $\pm$ 0.10 $\pm$ 0.22   \\  
$10.0-12.0$   & 1.93 $\pm$ 0.33 $\pm$ 0.10 $\pm$ 0.11 $\pm$ 0.24   \\ 
$12.0-14.0$   & 3.48 $\pm$ 0.64 $\pm$ 0.23 $\pm$ 0.20 $\pm$ 0.43 \\
\end{tabular} 
\end{table}
\begin{table}[h]
\centering
\caption{Differential \etac prompt production cross-section.}
\label{tab:PromptTable}
\begin{tabular}{c|
c @{\hspace{0.5\tabcolsep}} 
c @{\hspace{0.5\tabcolsep}}
c @{\hspace{0.5\tabcolsep}}
c @{\hspace{0.5\tabcolsep}}
c @{\hspace{0.5\tabcolsep}}
c @{\hspace{0.5\tabcolsep}}
c @{\hspace{0.5\tabcolsep}}
c @{\hspace{0.5\tabcolsep}}
c @{\hspace{0.5\tabcolsep}}
c @{\hspace{0.5\tabcolsep}}
c @{\hspace{0.5\tabcolsep}}
c}
\centering
\small
\pt  $[\gev\,]$     & \multicolumn{9}{c}{$d\sigma^{\text{prompt}}_{\etac}/d\pt$ $[\nb/\gev\,]$}  \\ \hline
$\phantom{1}6.5-\phantom{1}8.0$     & 488 &$\pm$& 111 &$\pm$& 17 &$\pm$& 28 &$\pm$& 64  \\ 
\phantom{1}$8.0-10.0$    & 157 &$\pm$& \phantom{1}25 &$\pm$& \phantom{1}6 &$\pm$& \phantom{1}9 &$\pm$& 22 \\  
$10.0-12.0$   & \phantom{1}63  &$\pm$& \phantom{1}11 &$\pm$& \phantom{1}3 &$\pm$& \phantom{1}4 &$\pm$& \phantom{1}9 \\ 
$12.0-14.0$   & \phantom{1}44  &$\pm$& \phantom{11}8 &$\pm$& \phantom{1}3 &$\pm$& \phantom{1}3 &$\pm$& \phantom{1}6 \\ 
\end{tabular} 
\end{table} 

\clearpage
\subsection{Production of \pmb{\etac} mesons from \pmb{\bquark}-hadron decays}
The results on relative \pt-differential \etac production in inclusive \bquark-hadron decays 
are shown in Table~\ref{tab:relSecondaryTable}. 
The results on \pt-differential \etac production cross-section in inclusive \bquark-hadron decays 
are shown in Table~\ref{tab:SecondaryTable}. As above, the last uncertainty includes the uncertainty on the \jpsi production cross-section.
\begin{table}[h]
\centering
\small
\caption{Relative \pt-differential \etac production cross-section in inclusive \bquark-hadron decays.}
\label{tab:relSecondaryTable}
\begin{tabular}{c|c}
\pt $[\gev\,]$     & $d\sigma^{\bquark}_{\etac}/d\sigma^{\bquark}_{\jpsi}$ \\ \hline 
$\phantom{1}6.5-\phantom{1}8.0$ & 0.41 $\pm$ 0.06 $\pm$ 0.01 $\pm$ 0.02 $\pm$ 0.05 \\ 
\phantom{1}$8.0 - 10.0$           & 0.61 $\pm$ 0.05 $\pm$ 0.03 $\pm$ 0.03 $\pm$ 0.08 \\  
$10.0 - 12.0$   & 0.45 $\pm$ 0.06 $\pm$ 0.02 $\pm$ 0.02 $\pm$ 0.06  \\ 
$12.0 - 14.0$   & 0.54 $\pm$ 0.07 $\pm$ 0.03 $\pm$ 0.02 $\pm$ 0.07  \\   
\end{tabular}
\end{table} 

\begin{table}[h]
\centering
\caption{The \pt-differential \etac production cross-section in inclusive \bquark-hadron decays.}
\label{tab:SecondaryTable}
\begin{tabular}{c|
c @{\hspace{0.5\tabcolsep}} 
c @{\hspace{0.5\tabcolsep}}
c @{\hspace{0.5\tabcolsep}}
c @{\hspace{0.5\tabcolsep}}
c @{\hspace{0.5\tabcolsep}}
c @{\hspace{0.5\tabcolsep}}
c @{\hspace{0.5\tabcolsep}}
c @{\hspace{0.5\tabcolsep}}
c @{\hspace{0.5\tabcolsep}}
c @{\hspace{0.5\tabcolsep}}
c @{\hspace{0.5\tabcolsep}}
c}
\centering
\small
\pt $[\gev\,]$     & \multicolumn{9}{c}{$d\sigma^{\bquark}_{\etac}/d\pt$ $[\nb/\gev\,]$}     \\ \hline 
$\phantom{1}6.5-\phantom{1}8.0$     
& 27.2 &$\pm$& 4.2 &$\pm$& 1.0 &$\pm$& 1.3 &$\pm$& 3.7 \\ 
\phantom{1}$8.0 - 10.0$    
& 18.8 &$\pm$& 1.5 &$\pm$& 0.8 &$\pm$& 0.9 &$\pm$& 2.6 \\  
$10.0 - 12.0$   
& \phantom{1}6.6 &$\pm$& 0.8 &$\pm$& 0.3 &$\pm$& 0.3 &$\pm$& 0.9 \\ 
$12.0 - 14.0$   
& \phantom{1}3.8 &$\pm$& 0.5 &$\pm$& 0.2 &$\pm$& 0.2 &$\pm$& 0.6 \\  
\end{tabular} 
\end{table}

%% file: LHCb_Authorship_flat_11-Jun-2019.tex
\centerline{\large\bf LHCb collaboration}
\begin{flushleft}
\small
R.~Aaij$^{28}$,
C.~Abell{\'a}n~Beteta$^{46}$,
T.~Ackernley$^{56}$,
B.~Adeva$^{43}$,
M.~Adinolfi$^{50}$,
H.~Afsharnia$^{6}$,
C.A.~Aidala$^{77}$,
S.~Aiola$^{22}$,
Z.~Ajaltouni$^{6}$,
S.~Akar$^{61}$
% ...

{\footnotesize \it
% ...
$ ^{6}$Universit{\'e} Clermont Auvergne, CNRS/IN2P3, LPC, Clermont-Ferrand, France\\
% ...
$ ^{22}$INFN Sezione di Milano, Milano, Italy\\
% ...
$ ^{28}$Nikhef National Institute for Subatomic Physics, Amsterdam, Netherlands\\
% ...
$ ^{43}$Instituto Galego de F{\'\i}sica de Altas Enerx{\'\i}as (IGFAE), Universidade de Santiago de Compostela, Santiago de Compostela, Spain\\
% ...
$ ^{46}$Physik-Institut, Universit{\"a}t Z{\"u}rich, Z{\"u}rich, Switzerland\\
% ...
$ ^{50}$H.H. Wills Physics Laboratory, University of Bristol, Bristol, United Kingdom\\
% ...
$ ^{56}$Oliver Lodge Laboratory, University of Liverpool, Liverpool, United Kingdom\\
% ...
$ ^{61}$University of Cincinnati, Cincinnati, OH, United States\\
% ...
$ ^{63}$Syracuse University, Syracuse, NY, United States\\
% ...
$ ^{77}$University of Michigan, Ann Arbor, United States, associated to $^{63}$\\
% ...
}
\end{flushleft}